\documentclass[lettersize,journal]{IEEEtran}
\IEEEoverridecommandlockouts
\usepackage{orcidlink}
\usepackage{amsfonts}
\usepackage{bbding}
\usepackage{pifont} 
\usepackage{array}
\usepackage{booktabs}

\usepackage{tikz, pgfplots}
 \pgfplotsset{compat=1.18}

\usepackage{textcomp}
\usepackage{stfloats}
\usepackage{url}
\usepackage{verbatim}
\usepackage{cite}
\usepackage{graphicx}
\usepackage{xcolor}
\usepackage{subfigure}
\usepackage{algorithm}
\usepackage{algpseudocode}
\usepackage{amsmath}
\usepackage{amssymb}
\usepackage{hyperref}
\usepackage{overpic, scalerel}
\usepackage{amsthm}

\hypersetup{
    colorlinks=true,     
    linkcolor=blue,      
    citecolor=blue,      
    urlcolor=blue,       
    pdfborder={0 0 0}    
}
\usepackage{orcidlink}

\usepackage{soul}

\usepackage{cite}

\title{Chance-Constrained Secrecy Optimization in Hybrid RIS-Empowered and UAV-Assisted Networks}
\author{Elhadj Moustapha Diallo$^{_{\orcidlink{0009-0000-4860-6253}}}$, Mamadou Aliou Diallo$^{_{\orcidlink{0009-0009-6860-8841}}}$, Abusaeed B. M. Adam, Muhammad Naeem Shah$^{_{\orcidlink{0000-0002-8966-1547}}}$
\thanks{E. M. Diallo and M. N. Shah are with Ningbo Ciruan Software Development Co., Ltd., Ningbo 315000 and Beijing SQUID Quantum Technology Co., Ltd., Beijing 100083, China.}
\thanks{M. A. Diallo is with School of Communications and Information Engineering, Chongqing University of Posts and Telecommunications, Chongqing, 400065, China.}
\thanks{A. B. M. is with the School of Science, Sudan University of Science and Technology, Khartoum, Sudan.}
}

\begin{document}
\pagenumbering{arabic}
\maketitle

\begin{abstract}
This paper considers a hybrid reconfigurable environment comprising a UAV-mounted reflecting RIS, an outdoor STAR-RIS enabling simultaneous transmission and reflection, and an indoor holographic RIS (H-RIS), jointly enhancing secure downlink communication for indoor and outdoor users. The system operates under user mobility, dynamic blockages, colluding idle and active eavesdroppers, and transceiver and surface hardware impairments. A 3GPP/ITU-compliant stochastic channel model is developed, capturing mobility-induced covariance evolution, outdoor–indoor penetration losses, and distortion-aware noise due to practical EVM-based impairments.
We aim to minimize the aggregate secrecy-outage probability subject to secrecy-rate constraints, QoS requirements, power limitations, and statistical CSI uncertainty. The resulting problem contains coupled secrecy and QoS chance constraints and nonlinear interactions among the BS beamforming vectors, multi-surface phase coefficients, and UAV position. To handle these difficulties, we derive rigorous Bernstein-type deterministic approximations for all chance constraints, yielding a distributionally robust reformulation. Building on this, we propose an alternating optimization framework that employs successive convex approximation (SCA) to convexify each block and solve the BS beamforming, RIS/STAR-RIS/H-RIS configuration, and UAV placement subproblems efficiently.
The proposed algorithm is shown to monotonically decrease a smooth surrogate of the secrecy-outage cost and converge to a stationary point of the robustified problem. Simulations based on 3GPP TR~38.901, TR~36.873, and ITU-R P.2109 demonstrate that integrating UAV-RIS, STAR-RIS, and H-RIS significantly reduces secrecy-outage probability compared with benchmark schemes and provides strong robustness to channel uncertainty, blockages, colluding eavesdroppers, and hardware impairments.
\end{abstract}
\section{Introduction}
Reconfigurable intelligent surfaces (RISs) have emerged as a key enabling technology for controlling the wireless propagation environment through
low-cost, nearly passive, and programmable manipulation of electromagnetic waves. By coherently adjusting the phase (and possibly amplitude) responses of large numbers of sub-wavelength elements, RISs can establish favorable virtual links, enhance coverage, and suppress interference without relying on active radio-frequency chains. Conventional reflecting RISs have demonstrated substantial gains in outdoor scenarios; however, their effectiveness degrades in heterogeneous indoor-outdoor environments due to penetration losses, frequent non-line-of-sight (NLoS) propagation, and dynamic blockages, which are particularly severe at millimeter-wave (mmWave) frequencies \cite{Kakati2025HybridRISMultiHop,Adam2023JitteringUAVHW}.

To overcome these limitations, advanced surface architectures such as simultaneously transmitting and reflecting RISs (STAR-RISs) and holographic
RISs (H-RISs) have been proposed. STAR-RISs enable concurrent signal reflection and transmission to serve users on both sides of the surface, thereby
facilitating seamless indoor-outdoor coverage and reliability enhancement \cite{Samy2024STARRISOutage}. H-RISs further push spatial controllability by approximating continuous electromagnetic apertures, enabling ultra-fine wavefront shaping and improved performance in dense multiuser settings
\cite{Ouamri2025DSPG_HRIS}. In parallel, unmanned aerial vehicles (UAVs) have been widely recognized as agile aerial platforms capable of exploiting 3D placement and mobility to establish advantageous line-of-sight (LoS) connections and provide adaptive coverage in dynamically changing environments. The integration of UAVs with RIS technology has therefore attracted growing interest as a promising solution for coverage extension and performance enhancement in next-generation wireless networks \cite{Adam2023JitteringUAVHW,Adam2024SecureUAVRIS}.

Despite these advances, guaranteeing physical-layer security in hybrid UAV-RIS/STAR-RIS/H-RIS architectures remains highly challenging. First, user
distributions are inherently heterogeneous: outdoor users may benefit from UAV-enabled LoS links and UAV-RIS or BS--RIS paths, while indoor users suffer from additional penetration attenuation and often require STAR-RIS or H-RIS assistance \cite{Kakati2025HybridRISMultiHop}. Second, UAV mobility, three-dimensional geometry variations, and human or vehicular blockages induce rapid statistical fluctuations in the channel, complicating secrecy-aware beamforming and surface configuration under realistic observation and estimation conditions. Third, practical systems are subject to hardware imperfections, including phase noise, nonlinear transceiver distortions, and finite-resolution RIS elements, which distort the effective cascaded channels and can significantly degrade achievable secrecy rates if ignored \cite{Adam2023JitteringUAVHW}. Finally, the adversarial landscape is evolving: eavesdroppers may be spatially distributed, fully colluding, and even active, injecting jamming signals to intentionally increase secrecy outage \cite{Diallo2025SecureContentFetching,Adam2023EnhancingSecrecyUAVRSMA}.

Most existing studies on RIS-aided physical-layer security still focus on single-surface deployments, simplified propagation assumptions, or idealized CSI models. Secure UAV-RIS systems under coordinated aerial and ground eavesdropping have been investigated using learning-based approaches
\cite{Diallo2025SecureContentFetching}, while hybrid indoor-outdoor mmWave secure communication has been addressed via multi-hop RIS or UAV-assisted
architectures \cite{Kakati2025HybridRISMultiHop}. However, these works do not capture the joint interaction of UAV-mounted RISs, STAR-RISs, and H RISs within a unified framework. Studies on H-RIS–aided systems primarily emphasize statistical quality-of-service (QoS) provisioning rather than secrecy-outage guarantees under colluding eavesdroppers \cite{Ouamri2025DSPG_HRIS}. Moreover, UAV-enabled secure communications based on rate-splitting multiple access (RSMA), deep unfolding, or reinforcement learning typically address only subsets of the above challenges and do not jointly incorporate multi-surface reconfigurability, statistical CSI uncertainty, and hardware impairments \cite{Adam2023EnhancingSecrecyUAVRSMA,Ouamri2025QuantumDRLUAVRSMA}.

Recently, more sophisticated learning-based and generative frameworks have been proposed to cope with the high-dimensional and nonconvex nature of UAV-and RIS-enabled network optimization, including multi-modal deep reinforcement learning, generative transformer-based primal--dual methods, and hierarchical DRL architectures \cite{Diallo2025MultiModalDRLUAVRIS,Adam2025MultiScaleGenTransformer,Adam2025GAI_HDRL_THz}. In parallel, real-time and security-aware precoding under RIS-assisted multiuser settings has been explored from a statistical and robustness perspective \cite{Adam2023RealtimeSecurityAwareRIS}. While promising, these approaches do not explicitly address secrecy-outage guarantees in hybrid
UAV-assisted RIS/STAR-RIS/H-RIS systems under colluding and potentially active eavesdroppers.

To bridge these gaps, this paper develops a comprehensive secure communication framework for a hybrid UAV-assisted RIS, STAR-RIS, and holographic RIS system serving mixed indoor and outdoor users. The considered network explicitly captures UAV mobility, dynamic blockages, outdoor–indoor propagation losses, transceiver and surface hardware impairments, and statistical CSI uncertainty, in a manner consistent with 3GPP and ITU-R standardized channel models.

Within this setting, we formulate a secrecy-outage minimization problem that jointly optimizes the BS beamforming vectors, the configurations of the UAV-mounted RIS, STAR-RIS, and H-RIS, as well as the UAV placement, subject to probabilistic secrecy-rate and QoS constraints and hardware-aware power limitations. The resulting design problem is particularly challenging due to the presence of secrecy and QoS chance constraints, which are highly nonconvex and tightly coupled through multi-surface cascaded channels, UAV positioning, and statistical channel uncertainties.

The key contributions of this paper are summarized as follows.
\begin{itemize}
    \item We propose a novel hybrid reconfigurable wireless architecture that jointly integrates a UAV-mounted reflecting RIS, an outdoor STAR-RIS, and an indoor holographic RIS to support secure downlink transmission for mixed indoor and outdoor users. Unlike existing works that consider single-surface or homogeneous deployments, the proposed architecture enables complementary manipulation of line-of-sight, penetration, and fine-grained indoor propagation paths within a unified framework.

    \item We develop a realistic stochastic channel and signal model that jointly accounts for UAV mobility, dynamic blockages, outdoor–indoor penetration losses, statistical CSI uncertainty, and transceiver as well as surface hardware impairments. The model is explicitly aligned with 3GPP TR 38.901, TR 36.873, and ITU-R P.2109, and incorporates EVM-based distortion noise, making it suitable for performance evaluation under practical operating conditions.

    \item We formulate a secrecy-outage–aware joint optimization problem that minimizes the weighted secrecy-outage probability subject to probabilistic secrecy-rate and QoS constraints, BS power limitations, and reconfigurable surface feasibility constraints. The problem jointly optimizes BS beamforming, UAV placement, and the configurations of the UAV-RIS, STAR-RIS, and H-RIS under fully colluding (and potentially active) eavesdroppers and statistical CSI uncertainty.

    \item To tackle the resulting nonconvex and distribution-dependent chance constraints, we derive Bernstein-type deterministic approximations for both secrecy and QoS outage constraints, yielding a distributionally robust reformulation that guarantees constraint satisfaction for all channel distributions consistent with given second-order statistics. This reformulation converts the original stochastic program into a tractable deterministic optimization problem.

    \item  We design an alternating optimization framework based on successive convex approximation (SCA) to efficiently solve the robustified problem. The proposed algorithm is shown to monotonically decrease a smooth surrogate of the secrecy-outage cost and converge to a stationary point. Numerical results under standardized channel models demonstrate that the proposed hybrid design achieves substantial secrecy-outage reductions compared with UAV-only, RIS-only, and STAR-RIS-only baselines, while maintaining reliable performance across a wide range of transmit powers and secrecy/QoS requirements.
\end{itemize}

\section{System Model}
We consider a downlink multiuser communication system in which a multi-antenna base station (BS) serves both outdoor and indoor legitimate users with the assistance of multiple reconfigurable intelligent surfaces (RISs). The system integrates three complementary reconfigurable structures: (i) a conventional reflecting RIS mounted on an unmanned aerial vehicle (UAV), (ii) a simultaneous transmitting and reflecting RIS (STAR-RIS) deployed on the exterior wall of a building, and (iii) an indoor holographic RIS (H-RIS) providing fine-grained electromagnetic wavefront manipulation inside the building. This hybrid architecture enables coverage extension, outdoor-to-indoor signal penetration, and enhanced spatial controllability of the wireless propagation environment.

Let $\mathcal{K}_{\mathrm{out}}$ and $\mathcal{K}_{\mathrm{in}}$ denote the sets of outdoor and indoor legitimate users, respectively, and define
$\mathcal{K}=\mathcal{K}_{\mathrm{out}}\cup\mathcal{K}_{\mathrm{in}}$. The system is subject to security threats from a set of idle and active eavesdroppers, denoted by $\mathcal{E}=\mathcal{E}_{\mathrm{idle}}\cup\mathcal{E}_{\mathrm{act}}$, which attempt to wiretap or actively disrupt the communication links. All eavesdroppers are assumed to be fully colluding and capable of jointly processing their received signals.

\subsection{Network Architecture and Transmission Model}
The BS is equipped with $N_t$ antennas and employs linear precoding to simultaneously transmit independent information streams to all legitimate users. The transmitted signal is given by
\begin{equation}
\mathbf{x}=\sum_{k\in\mathcal{K}} \mathbf{w}_k s_k,
\label{eq:bs_signal}
\end{equation}
where $\mathbf{w}_k\in\mathbb{C}^{N_t}$ denotes the precoding vector associated with user $k$, and $s_k$ is the corresponding information symbol with unit average power, i.e., $\mathbb{E}[|s_k|^2]=1$. The total transmit power at the BS is constrained by
\begin{equation}
\sum_{k\in\mathcal{K}}\|\mathbf{w}_k\|^2 \le P_{\max}.
\label{eq:power_constraint}
\end{equation}
This formulation captures a general multiuser downlink transmission model and encompasses a wide range of linear precoding strategies.

The BS is deployed outdoors at height $h_{\mathrm{BS}}$. A UAV carrying a passive RIS with $M_U$ reflecting elements hovers at altitude $h_{\mathrm{UAV}}$, with horizontal position $\mathbf{p}_U=(x_U,y_U)$, which may be optimized within a feasible region $\mathcal{P}_U$. A STAR-RIS comprising $M_S$ controllable elements is mounted on the exterior façade of a building, while an H-RIS with $M_H$ controllable coefficients is deployed on an interior wall.

Outdoor legitimate users are spatially distributed according to a homogeneous Poisson point process (PPP) with density $\lambda_{\mathrm{out}}$ over a circular region of radius $R_{\mathrm{out}}$. Indoor users follow an independent PPP with density $\lambda_{\mathrm{in}}$ within the building footprint. Colluding eavesdroppers are independently distributed according to a PPP with density $\lambda_{\mathrm{eve}}$. These spatial models enable statistical performance evaluation under random network realizations.

\subsection{Reconfigurable Surface Modeling}
The UAV-mounted RIS operates in pure reflection mode and is modeled by a diagonal phase-shift matrix
\begin{equation}
\boldsymbol{\Theta}_U = \mathrm{diag}\!\left(e^{j\theta_{U,1}},\dots,e^{j\theta_{U,M_U}}\right),
\end{equation}
where $\theta_{U,m}\in[0,2\pi)$ denotes the phase shift applied by the $m$-th element. This unit-modulus constraint reflects the passive nature of the reflecting elements.

Each STAR-RIS element splits the incident signal into transmitted and reflected components. The corresponding complex coefficients are modeled as
\begin{equation}
t_m = \rho_m e^{j\theta_{T,m}},\qquad
r_m = \sqrt{1-\rho_m^2}\, e^{j\theta_{R,m}},
\end{equation}
where $0\le\rho_m\le 1$ enforces per-element energy conservation. The associated transmission and reflection matrices are given by
\begin{equation}
\boldsymbol{\Theta}_T = \mathrm{diag}(t_1,\dots,t_{M_S}),\qquad
\boldsymbol{\Theta}_R = \mathrm{diag}(r_1,\dots,r_{M_S}).
\end{equation}
This model captures the simultaneous reflection and transmission capability of STAR-RIS elements.

The H-RIS is modeled as a discretized continuous aperture
\begin{equation}
\boldsymbol{\Theta}_H = \mathrm{diag}\!\left(\alpha_{H,1}e^{j\theta_{H,1}},\dots,\alpha_{H,M_H}e^{j\theta_{H,M_H}}\right),
\end{equation}
where $\alpha_{H,m}\le\alpha_{\max}$ captures physically realizable amplitude constraints imposed by metamaterial-based implementations.

To accommodate different hardware realizations, the coefficients of each surface are constrained to lie in individual feasibility sets
\begin{equation}
\boldsymbol{\Theta}_U\in\mathcal{F}_U,\quad
(\boldsymbol{\Theta}_T,\boldsymbol{\Theta}_R)\in\mathcal{F}_S,\quad
\boldsymbol{\Theta}_H\in\mathcal{F}_H,
\end{equation}
which will be explicitly enforced in the subsequent optimization framework.

\subsection{Propagation and Channel Model}
All wireless links are modeled in accordance with the 3GPP channel modeling framework specified in TR~38.901 and TR~36.873, depending on whether the
propagation environment corresponds to outdoor, indoor, or outdoor-to-indoor scenarios~\cite{3gpp38901,3gpp36873}. These models capture pathloss, shadowing, line-of-sight (LoS) probability, and small-scale fading characteristics for Urban Micro (UMi) street-canyon and Indoor Hotspot (InH) environments.
\begin{equation}
\mathbf{h}_{a,b} = \sqrt{\beta_{a,b}(d_{a,b},\mathcal{S}_{a,b})}\,\mathbf{g}_{a,b},
\label{eq:channel_def}
\end{equation}
where $d_{a,b}$ denotes the link distance and $\mathcal{S}_{a,b}$ represents the propagation state capturing line-of-sight (LoS), non-line-of-sight (NLoS), and blockage conditions. The large-scale fading coefficient $\beta_{a,b}(\cdot)$ incorporates distance-dependent pathloss and log-normal shadowing, while $\mathbf{g}_{a,b}$ denotes the small-scale fading component, following Rician or Rayleigh distributions depending on the LoS condition.

Outdoor-to-indoor penetration losses are incorporated into the large-scale fading coefficient $\beta_{a,b}(\cdot)$ according to the ITU-R P.2109
multilayer penetration loss model, which accounts for frequency-dependent attenuation and material properties such as concrete and low-emissivity
glass~\cite{iturp2109}.

User mobility and environmental dynamics induce time variations in both $d_{a,b}(t)$ and $\mathcal{S}_{a,b}(t)$. Indoor users follow a random waypoint mobility model with speeds uniformly distributed in $[0.5,1.0]$ m/s, while the UAV follows a three-dimensional kinematic model with bounded velocity $v_{\mathrm{UAV}}$ and fixed altitude $h_{\mathrm{UAV}}$. Dynamic blockages caused by humans, vehicles, and UAV motion are modeled using a Markov process with empirically motivated blockage durations and depths.

\subsection{Signal Model with Hardware Impairments}
Due to non-ideal radio-frequency chains, the transmitted signal at the BS is subject to residual hardware impairments. Specifically, the actual transmitted
waveform is given by
\begin{equation}
\mathbf{x}_{\mathrm{tx}} = \mathbf{x} + \boldsymbol{\eta}_t,
\label{eq:tx_distortion}
\end{equation}
where $\boldsymbol{\eta}_t$ denotes transmitter distortion noise. Its covariance
is modeled as $\mathbb{E}[\boldsymbol{\eta}_t\boldsymbol{\eta}_t^H] =\kappa_t \mathrm{diag}(\mathbb{E}[\mathbf{x}\mathbf{x}^H]),$ with $\kappa_t$ denoting the transmitter error vector magnitude (EVM).

At the receiver side, each legitimate user and eavesdropper experiences additional distortion due to receiver hardware impairments. The received signal at a generic receiver $u$ is expressed as
\begin{equation}
y_u = y_u^{\mathrm{ideal}} + \eta_{r,u},
\label{eq:rx_distortion}
\end{equation}
where $\eta_{r,u}$ denotes receiver distortion noise whose variance scales with the instantaneous received signal power through the receiver EVM parameter $\kappa_{r,u}$.

\subsection{CSI Uncertainty, SINR, and Secrecy Metrics}
The BS operates with imperfect channel state information (CSI). For each link $a\rightarrow b$, the available CSI is modeled as
\begin{equation}
\mathbf{h}_{a,b} = \widehat{\mathbf{h}}_{a,b} + \Delta\mathbf{h}_{a,b},
\label{eq:error_model}
\end{equation}
where $\widehat{\mathbf{h}}_{a,b}$ denotes the channel estimate available at the BS and $\Delta\mathbf{h}_{a,b}$ represents the channel estimation error. The error term captures the impact of mobility, blockage dynamics, and estimation imperfections, and is assumed to satisfy $\mathbb{E}[\|\Delta\mathbf{h}_{a,b}\|^2]\le\sigma_{a,b}^2$, consistent with a second-order statistical uncertainty model.

By coherently combining the direct BS--user link and all RIS-assisted cascaded paths, the effective channel observed by legitimate user $k$ is expressed as
\begin{equation}
\mathbf{h}_{k}^{\mathrm{eff}} = \mathbf{h}_{k}^{\mathrm{dir}} + \mathbf{h}_{k}^{\mathrm{UAV}} + \mathbf{h}_{k}^{\mathrm{STAR-T}} +
\mathbf{h}_{k}^{\mathrm{STAR-R}} + \mathbf{h}_{k}^{\mathrm{Holo}},
\label{eq:effective_channel}
\end{equation}
which explicitly reflects the joint contribution of the UAV-mounted RIS, the outdoor STAR-RIS, and the indoor holographic RIS.

Based on the impaired signal model and the effective channel in \eqref{eq:effective_channel}, the signal-to-interference-plus-noise ratio (SINR) at legitimate user $k$ is given by
\begin{equation}
\gamma_k^{(\ell)} = \frac{ |\mathbf{h}_{k}^{\mathrm{eff}H}\mathbf{w}_k|^2}{\sum_{i\neq k}|\mathbf{h}_{k}^{\mathrm{eff}H}\mathbf{w}_i|^2
+ \sigma_k^2 + \sigma_{k,\mathrm{dist}}^2 + I_{k,\mathrm{jam}}},
\label{eq:SINR_legit}
\end{equation}
where $\sigma_k^2$ denotes the thermal noise power, $\sigma_{k,\mathrm{dist}}^2$ accounts for receiver-side hardware distortion, and $I_{k,\mathrm{jam}}$ represents the interference induced by active jamming eavesdroppers when present. The corresponding instantaneous communication rate is
\begin{equation}
R_k^{(\ell)}=\log_2\!\left(1+\gamma_k^{(\ell)}\right).
\end{equation}

Each eavesdropper observes a superposition of all transmitted signals corrupted by thermal noise and hardware impairments. The SINR at eavesdropper $e$ when attempting to decode the signal intended for user $k$ is given by
\begin{equation}
\gamma_{k,e} = \frac{ |\mathbf{h}_{e}^{\mathrm{eff}H}\mathbf{w}_k|^2 }{ \sum_{i\neq k}|\mathbf{h}_{e}^{\mathrm{eff}H}\mathbf{w}_i|^2
+ \sigma_e^2 + \sigma_{e,\mathrm{dist}}^2}.
\end{equation}

We assume that all eavesdroppers are fully colluding and capable of jointly processing their received signals. Under this worst-case assumption, the aggregate wiretap SINR for user $k$ is modeled as
\begin{equation}
\gamma_k^{(e)}=\sum_{e\in\mathcal{E}}\gamma_{k,e},
\label{eq:colluding_sinr}
\end{equation}
which corresponds to coherent combining of the individual eavesdropping links and provides a conservative characterization of the wiretap capability. The resulting wiretap rate is $R_k^{(e)}=\log_2(1+\gamma_k^{(e)})$. Accordingly, the instantaneous secrecy rate of user $k$ is defined as
\begin{equation}
R_k^{\mathrm{sec}} = \big[ R_k^{(\ell)}-R_k^{(e)} \big]^+.
\label{eq:secrecy_rate}
\end{equation}

Let $R_k^{\mathrm{QoS}}$ and $R_k^{\mathrm{sec,min}}$ denote the minimum communication and secrecy rate requirements, respectively. A secrecy outage event is declared whenever either the secrecy constraint or the QoS constraint is violated, reflecting the fact that secure communication is only meaningful when a minimum service quality is guaranteed. The resulting secrecy-outage probability is defined as
\begin{equation}
P_k^{\mathrm{out}} = \Pr\!\left\{ R_k^{\mathrm{sec}}<R_k^{\mathrm{sec,min}} \;\text{or}\; R_k^{(\ell)}<R_k^{\mathrm{QoS}} \right\},
\label{eq:outage_prob}
\end{equation}
which serves as the key performance metric optimized in the subsequent section.

\section{Problem Formulation}
In this section, we translate the secrecy, QoS, and reliability requirements introduced in Section~III into a rigorous optimization framework. The objective is to design the BS precoders, the configurations of the UAV-RIS, STAR-RIS, and H-RIS, as well as the UAV position, so as to minimize the secrecy outage experienced by the legitimate users under power, hardware, and reliability constraints. All outage metrics are defined with respect to the underlying distributions induced by user mobility, dynamic blockages, channel-estimation errors, and hardware impairments.
\subsection{Design Variables and Feasible Sets}
The optimization variables are collected in the set
\begin{equation}
    \boldsymbol{\Xi}
    =
    \Big\{
        \{\mathbf{w}_k\}_{k\in\mathcal{K}},
        \boldsymbol{\Theta}_U,
        \boldsymbol{\Theta}_T,
        \boldsymbol{\Theta}_R,
        \boldsymbol{\Theta}_H,
        \mathbf{p}_U
    \Big\}.
    \label{eq:design_variables}
\end{equation}

The beamforming vectors $\{\mathbf{w}_k\}$ must satisfy the BS power budget in \eqref{eq:power_constraint}. The UAV-RIS reflection matrix $\boldsymbol{\Theta}_U$, the STAR-RIS transmission and reflection matrices $(\boldsymbol{\Theta}_T,\boldsymbol{\Theta}_R)$, and the holographic-RIS matrix $\boldsymbol{\Theta}_H$ are constrained to lie in their respective feasible sets
\begin{equation}
    \boldsymbol{\Theta}_U \in \mathcal{F}_U,\quad
    (\boldsymbol{\Theta}_T,\boldsymbol{\Theta}_R) \in \mathcal{F}_S,\quad
    \boldsymbol{\Theta}_H \in \mathcal{F}_H,
    \label{eq:RIS_feasible_sets}
\end{equation}
which capture amplitude and phase constraints, phase quantization, and power conservation at each surface. The UAV position $\mathbf{p}_U$ is constrained to a safe 3D operating region
\begin{equation}
    \mathbf{p}_U \in \mathcal{P}_U,
    \label{eq:UAV_region}
\end{equation}
accounting for altitude limits, no-fly zones, and mission geometry. The statistical channel-error model in \eqref{eq:error_model} is implicitly imposed via the bounds
\begin{equation}
    \mathbb{E}\!\left[\big\|\Delta\mathbf{h}_{a,b}\big\|^2\right] \le \sigma_{a,b}^2,\quad \forall (a,b),
    \label{eq:channel_error_bound}
\end{equation}
which determine the distribution of the random rates and, consequently, of the outage events.

\subsection{Secrecy-Outage Based Objective}
For each user $k\in\mathcal{K}$, the combined outage event $\mathcal{F}_k$ and the corresponding outage probability $P_k^{\mathrm{out}}$ are defined in \eqref{eq:outage_prob}. These probabilities depend on the design variables $\boldsymbol{\Xi}$ through the SINR expressions in \eqref{eq:SINR_legit} and \eqref{eq:colluding_sinr}, the secrecy rate in \eqref{eq:secrecy_rate}, and the underlying random channel realizations induced by user mobility, dynamic blockages, and hardware impairments. To explicitly highlight this dependence, we write
\begin{equation}
    P_k^{\mathrm{out}}(\boldsymbol{\Xi})
    =
    \Pr\!\left\{
        R_k^{\mathrm{sec}}(\boldsymbol{\Xi}) < R_k^{\mathrm{sec,min}}
        \ \text{or}\
        R_k^{(\ell)}(\boldsymbol{\Xi}) < R_k^{\mathrm{QoS}}
    \right\}.
    \label{eq:outage_prob_Xi}
\end{equation}

Unlike average secrecy-rate maximization, the secrecy-outage metric explicitly captures the reliability of secure communication under stochastic channel variations and uncertainty, making it particularly suitable for mobility-aware and blockage-prone environments. To balance the outage performance across heterogeneous users, we introduce non-negative weighting coefficients $\{\omega_k\}_{k\in\mathcal{K}}$ and define the network-level secrecy-outage cost as
\begin{equation}
    \Phi(\boldsymbol{\Xi})
    =
    \sum_{k\in\mathcal{K}} \omega_k
    P_k^{\mathrm{out}}(\boldsymbol{\Xi}).
    \label{eq:global_cost}
\end{equation}

The primary design objective is to minimize $\Phi(\boldsymbol{\Xi})$ subject to the physical, reliability, and security constraints of the system.

\subsection{Chance-Constrained Reliability and Secrecy Requirements}
In addition to minimizing the aggregate secrecy-outage cost, it is often desirable to enforce explicit per-user reliability and secrecy guarantees. In highly dynamic environments with imperfect CSI, deterministic instantaneous constraints are generally infeasible or overly conservative. Instead, probabilistic guarantees provide a more meaningful characterization of long-term service reliability and security.

Specifically, for each legitimate user $k$, we require that the secrecy rate and the QoS rate exceed their respective thresholds with high probability:
\begin{equation}
    \Pr\!\left\{
        R_k^{\mathrm{sec}}(\boldsymbol{\Xi}) \ge R_k^{\mathrm{sec,min}}
    \right\} \ge 1 - \epsilon_k,
    \label{eq:secrecy_chance_constraint}
\end{equation}
\begin{equation}
    \Pr\!\left\{
        R_k^{(\ell)}(\boldsymbol{\Xi}) \ge R_k^{\mathrm{QoS}}
    \right\} \ge 1 - \delta_k,
    \label{eq:QoS_chance_constraint}
\end{equation}
where $\epsilon_k$ and $\delta_k$ denote the maximum tolerable violation probabilities for secrecy and QoS, respectively. By De Morgan’s law, these chance constraints are equivalent to requiring that the individual secrecy-outage and QoS-outage probabilities satisfy
\begin{equation}
    \Pr\{\mathcal{O}_k\}
    \le \epsilon_k,\qquad
    \Pr\{\mathcal{Q}_k\}
    \le \delta_k,\quad \forall k\in\mathcal{K}.
\end{equation}

Furthermore, if desired, a direct upper bound on the combined secrecy-outage probability can be imposed as
\begin{equation}
    P_k^{\mathrm{out}}(\boldsymbol{\Xi}) \le \overline{P}_k,\quad \forall k\in\mathcal{K},
    \label{eq:Pk_out_upper_bound}
\end{equation}
where $\overline{P}_k$ specifies the maximum acceptable outage level for user $k$.

\subsection{Secrecy-Outage Minimization Problem}
By collecting the design variables in \eqref{eq:design_variables}, the secrecy-outage–aware joint optimization of BS beamforming, UAV-mounted RIS, STAR-RIS, holographic RIS, and UAV placement is formulated as the following constrained stochastic program. Let the weighted secrecy-outage cost be defined as
\begin{equation}
    \Phi(\boldsymbol{\Xi})
    = \sum_{k\in\mathcal{K}}\omega_k\, P_k^{\mathrm{out}}(\boldsymbol{\Xi}),
\end{equation}
where $P_k^{\mathrm{out}}(\boldsymbol{\Xi})$ is given in \eqref{eq:outage_prob_Xi}. The resulting secrecy-outage minimization problem is formulated as

\begin{subequations}
\label{eq:ProbP1}
\begin{align}
    \text{(P1)}\qquad
    \min_{\boldsymbol{\Xi}} \quad
    & \Phi(\boldsymbol{\Xi})
    \label{eq:P1_obj} \\[1mm]
    \text{s.t.}\quad
    & \sum_{k\in\mathcal{K}}\|\mathbf{w}_k\|^2 \le P_{\max},
        \label{eq:P1_power_constraint} \\[1mm]
    & \boldsymbol{\Theta}_U \in \mathcal{F}_U,\quad
      (\boldsymbol{\Theta}_T,\boldsymbol{\Theta}_R)\in\mathcal{F}_S,
      \boldsymbol{\Theta}_H \in \mathcal{F}_H,
        \label{eq:P1_RIS_constraints} \\[1mm]
    & \mathbf{p}_U \in \mathcal{P}_U,
        \label{eq:P1_UAV_constraint} \\[1mm]
    & \mathbb{E}\!\left[
        \|\Delta\mathbf{h}_{a,b}\|^2
      \right] \le \sigma_{a,b}^2, \forall (a,b),
        \label{eq:P1_error_constraint} \\[1mm]
    & \Pr\!\left\{
        R_k^{\mathrm{sec}}(\boldsymbol{\Xi})
        \ge R_k^{\mathrm{sec,min}}
      \right\}
      \ge 1-\epsilon_k, \forall k\in\mathcal{K},
        \label{eq:P1_secrecy_chance} \\[1mm]
    & \Pr\!\left\{
        R_k^{(\ell)}(\boldsymbol{\Xi})
        \ge R_k^{\mathrm{QoS}}
      \right\}
      \ge 1-\delta_k,\forall k\in\mathcal{K}.
        \label{eq:P1_QoS_chance}
\end{align}
\end{subequations}

Problem~\eqref{eq:ProbP1} is a highly non-convex and distribution-dependent optimization problem. The non-convexity arises from: (i) the coupled structure of the effective channels in \eqref{eq:effective_channel}, (ii) the multiplicative interaction between beamforming vectors and the RIS, STAR-RIS, and H-RIS coefficients, and (iii) the dependence of large-scale fading on the UAV position. In addition, the presence of probabilistic secrecy and QoS constraints in \eqref{eq:P1_secrecy_chance}--\eqref{eq:P1_QoS_chance} renders the problem intractable in its original stochastic form. These challenges motivate the development of tractable deterministic approximations, which are addressed in Section~IV.

\subsection{Alternative Max--Min Secrecy-Rate Formulation}
An alternative fairness-oriented design objective is to maximize the secrecy rate that can be simultaneously guaranteed (with high probability) to all users. Introducing an auxiliary variable $\tau$ representing the guaranteed secrecy throughput, we obtain

\begin{subequations}
\label{eq:ProbP2}
\begin{align}
    \text{(P2)}\qquad
    \max_{\boldsymbol{\Xi},\,\tau} \quad
    & \tau
    \label{eq:P2_obj} \\[1mm]
    \text{s.t.}\quad
    & \eqref{eq:P1_power_constraint},\ 
      \eqref{eq:P1_RIS_constraints},\ 
      \eqref{eq:P1_UAV_constraint},\ 
      \eqref{eq:P1_error_constraint},
      \label{eq:P2_InheritedConstraints} \\[1mm]
    & \Pr\!\left\{
        R_k^{\mathrm{sec}}(\boldsymbol{\Xi})
        \ge \tau
      \right\}
      \ge 1-\epsilon_k,\quad \forall k\in\mathcal{K},
      \label{eq:P2_secrecy_chance} \\[1mm]
    & \Pr\!\left\{
        R_k^{(\ell)}(\boldsymbol{\Xi})
        \ge R_k^{\mathrm{QoS}}
      \right\}
      \ge 1-\delta_k,\quad \forall k\in\mathcal{K}.
      \label{eq:P2_QoS_chance}
\end{align}
\end{subequations}

\section{Deterministic Reformulation of Chance Constraints}
\label{sec:deterministic_reformulation}
The secrecy-outage–aware design problem~(P1) in \eqref{eq:ProbP1} is fundamentally complicated by the probabilistic secrecy and QoS constraints in \eqref{eq:P1_secrecy_chance}--\eqref{eq:P1_QoS_chance}, which are defined with respect to random channel-estimation errors, user mobility, dynamic blockage processes, and hardware-induced distortions. These sources of uncertainty render the feasible set of~(P1) distribution-dependent and generally intractable to characterize in closed form.  

In this section, we develop conservative deterministic approximations of the secrecy and QoS chance constraints by exploiting a Bernstein-type inequality for quadratic forms of complex Gaussian random vectors. The resulting reformulation yields a distributionally robust deterministic problem that guarantees the prescribed reliability and secrecy levels under second-order statistical uncertainty. This reformulation enables the efficient application of a successive convex approximation (SCA) combined with alternating optimization (AO), which will be detailed in the subsequent section.

\subsection{Channel-Error Model and Quadratic Representation}
Recall from \eqref{eq:error_model} that the effective channel from the BS, through the cascaded RIS, STAR-RIS, and H-RIS links, to a generic receiver $u$ (legitimate user or eavesdropper) is modeled as
\begin{equation}
    \mathbf{h}_u
    =
    \widehat{\mathbf{h}}_u + \Delta\mathbf{h}_u,
    \quad
    \mathbb{E}\!\big[\Delta\mathbf{h}_u\big] = \mathbf{0},
    \quad
    \mathbb{E}\!\big[\Delta\mathbf{h}_u \Delta\mathbf{h}_u^H\big]
    = \mathbf{C}_u,
    \label{eq:channel_error_model_IV}
\end{equation}
where $\widehat{\mathbf{h}}_u$ denotes the estimated effective channel available at the BS and $\Delta\mathbf{h}_u \sim \mathcal{CN}(\mathbf{0},\mathbf{C}_u)$ represents the aggregate channel estimation error. The covariance matrix $\mathbf{C}_u$ captures second-order statistical uncertainty induced by user mobility, dynamic blockages, and estimation imperfections, and serves as the basis for the subsequent distributionally robust reformulation.

For notational brevity, the impact of hardware impairments is absorbed into an effective noise-plus-distortion term, and the SINR at legitimate user $k$ in \eqref{eq:SINR_legit} is equivalently written as
\begin{equation}
    \gamma_k^{(\ell)}
    =
    \frac{
        |\mathbf{h}_k^H \mathbf{w}_k|^2
    }{
        \sum_{i\neq k}|\mathbf{h}_k^H \mathbf{w}_i|^2 + \sigma_{k,\mathrm{eff}}^2
    },
    \label{eq:SINR_legit_compact}
\end{equation}
where $\sigma_{k,\mathrm{eff}}^2$ collects thermal noise, receiver distortion noise, and potential jamming interference. An analogous compact representation holds for the eavesdroppers.

Given a target SINR threshold $\gamma_k^{\mathrm{QoS}}$ associated with the QoS requirement, the constraint $\gamma_k^{(\ell)} \ge \gamma_k^{\mathrm{QoS}}$ can be equivalently expressed as
\begin{equation}
    |\mathbf{h}_k^H \mathbf{w}_k|^2
    \;\ge\;
    \gamma_k^{\mathrm{QoS}}
    \Big(
        \sum_{i\neq k}|\mathbf{h}_k^H \mathbf{w}_i|^2
        + \sigma_{k,\mathrm{eff}}^2
    \Big).
    \label{eq:SINR_QoS_constraint_raw}
\end{equation}
Substituting $\mathbf{h}_k = \widehat{\mathbf{h}}_k + \Delta\mathbf{h}_k$ into \eqref{eq:SINR_QoS_constraint_raw} and rearranging terms yields a quadratic inequality in the complex Gaussian error vector $\Delta\mathbf{h}_k$, namely
\begin{equation}
    \Delta\mathbf{h}_k^H \mathbf{A}_k^{\mathrm{QoS}}(\boldsymbol{\Xi}) \Delta\mathbf{h}_k
    + 2\Re\!\left\{\mathbf{b}_k^{\mathrm{QoS}}(\boldsymbol{\Xi})^H \Delta\mathbf{h}_k\right\}
    + c_k^{\mathrm{QoS}}(\boldsymbol{\Xi})
    \;\ge\; 0,
    \label{eq:quadratic_QoS_form}
\end{equation}
where
\begin{align}
    \mathbf{A}_k^{\mathrm{QoS}}
    &=
    \mathbf{w}_k\mathbf{w}_k^H
    - \gamma_k^{\mathrm{QoS}} \sum_{i\neq k}
        \mathbf{w}_i\mathbf{w}_i^H,
    \label{eq:A_k_QoS_def} \\
    \mathbf{b}_k^{\mathrm{QoS}}
    &=
    \mathbf{w}_k\mathbf{w}_k^H \widehat{\mathbf{h}}_k
    - \gamma_k^{\mathrm{QoS}}
      \sum_{i\neq k}\mathbf{w}_i\mathbf{w}_i^H \widehat{\mathbf{h}}_k,
    \label{eq:b_k_QoS_def} \\
    c_k^{\mathrm{QoS}}
    &=
    \widehat{\mathbf{h}}_k^H
    \mathbf{A}_k^{\mathrm{QoS}}
    \widehat{\mathbf{h}}_k
    - \gamma_k^{\mathrm{QoS}} \sigma_{k,\mathrm{eff}}^2.
    \label{eq:c_k_QoS_def}
\end{align}
For clarity of exposition, the explicit dependence of $\mathbf{A}_k^{\mathrm{QoS}}$, $\mathbf{b}_k^{\mathrm{QoS}}$, and $c_k^{\mathrm{QoS}}$ on the optimization variables $\boldsymbol{\Xi}$ is omitted when no ambiguity arises.

An analogous quadratic representation applies to the secrecy constraints. In particular, the secrecy-rate condition $R_k^{\mathrm{sec}} \ge R_k^{\mathrm{sec,min}}$ can be equivalently enforced by requiring $\gamma_k^{(\ell)} \ge \gamma_k^{\mathrm{sec},\ell}$ and $\gamma_k^{(e)} \le \gamma_k^{\mathrm{sec},e}$ for appropriately chosen thresholds $\gamma_k^{\mathrm{sec},\ell}$ and $\gamma_k^{\mathrm{sec},e}$. The resulting eavesdropper-side SINR constraint can be reformulated as a quadratic inequality of the form \eqref{eq:quadratic_QoS_form}, with problem-dependent matrices $(\mathbf{A}_k^{\mathrm{sec}}, \mathbf{b}_k^{\mathrm{sec}}, c_k^{\mathrm{sec}})$ constructed from the aggregated colluding eavesdropper channels.

\subsection{Bernstein-Type Deterministic Approximation}
We now recall a standard Bernstein-type inequality for quadratic forms of complex Gaussian random vectors, which has been widely adopted for conservative reformulation of chance-constrained optimization problems~\cite{ref:bernstein1,ref:bernstein2}.

\medskip
\noindent\textbf{Lemma 1 (Bernstein-type inequality):}
Let $\Delta\mathbf{h}\sim\mathcal{CN}(\mathbf{0},\mathbf{C})$ be a circularly symmetric complex Gaussian random vector with known covariance matrix $\mathbf{C}\succeq\mathbf{0}$, and consider the quadratic random variable
\begin{equation}
    Z
    =
    \Delta\mathbf{h}^H \mathbf{A} \Delta\mathbf{h}
    + 2\Re\!\{\mathbf{b}^H \Delta\mathbf{h}\}
    + c,
    \label{eq:Z_def}
\end{equation}
where $\mathbf{A}\in\mathbb{C}^{N\times N}$ is a Hermitian matrix, $\mathbf{b}\in\mathbb{C}^{N}$ is a deterministic vector, and $c\in\mathbb{R}$ is a deterministic scalar. For any prescribed violation probability $\epsilon\in(0,1)$, a sufficient condition for the chance constraint
\begin{equation}
    \Pr\{Z \ge 0\} \ge 1-\epsilon
    \label{eq:bernstein_target}
\end{equation}
to hold is the existence of a slack variable $\tau\ge 0$ such that
\begin{subequations}
\label{eq:bernstein_det}
\begin{align}
    & \mathrm{tr}(\mathbf{A}\mathbf{C}) + c
      - \tau \sqrt{2\ln(1/\epsilon)}
      \ge 0,
      \label{eq:bernstein_det_a} \\
    & \left\|
        \mathbf{C}^{1/2}\mathbf{A}\mathbf{C}^{1/2}
      \right\|_{\!F}
      \le \tau,
      \label{eq:bernstein_det_b} \\
    & \left\|
        \mathbf{C}^{1/2}\mathbf{b}
      \right\|_2
      \le \frac{\tau}{\sqrt{2}},
      \label{eq:bernstein_det_c}
\end{align}
\end{subequations}
where $\|\cdot\|_{F}$ and $\|\cdot\|_2$ denote the Frobenius and Euclidean norms, respectively.

The deterministic constraints in \eqref{eq:bernstein_det} are jointly convex in $(\mathbf{A},\mathbf{b},c)$ for fixed $\mathbf{C}$ and $\epsilon$, and can be equivalently expressed as second-order cone (SOC) constraints using standard transformations. As such, Lemma~1 provides a tractable and conservative deterministic approximation of the chance constraint \eqref{eq:bernstein_target}, guaranteeing constraint satisfaction for all realizations consistent with the specified second-order statistics.

\subsection{Deterministic QoS Chance Constraints}
We first apply Lemma~1 to the QoS chance constraints in \eqref{eq:P1_QoS_chance}. For a given user $k$, the QoS outage event $\mathcal{Q}_k=\{R_k^{(\ell)} < R_k^{\mathrm{QoS}}\}$ is equivalent to $\gamma_k^{(\ell)} < \gamma_k^{\mathrm{QoS}}$, which in turn corresponds to the violation of the quadratic inequality in \eqref{eq:quadratic_QoS_form}. Consequently, the QoS chance constraint \eqref{eq:P1_QoS_chance} can be equivalently rewritten as
\begin{align}
    \Pr\!\left\{
        \Delta\mathbf{h}_k^H \mathbf{A}_k^{\mathrm{QoS}} \Delta\mathbf{h}_k
        + 2\Re\!\left\{
            \mathbf{b}_k^{\mathrm{QoS}H} \Delta\mathbf{h}_k
        \right\}
        + c_k^{\mathrm{QoS}}
        \ge 0
    \right\}\nonumber\\
    \ge 1-\delta_k,
    \label{eq:QoS_bernstein_target}
\end{align}
which is in the same canonical form as the chance constraint in Lemma~1.

By directly invoking Lemma~1, a sufficient deterministic condition for satisfying \eqref{eq:QoS_bernstein_target} is the existence of a slack variable $\tau_k^{\mathrm{QoS}}\ge 0$ such that
\begin{subequations}
\label{eq:QoS_bernstein_det}
\begin{align}
    & \mathrm{tr}\!\left(
        \mathbf{A}_k^{\mathrm{QoS}}\mathbf{C}_k
      \right)
      + c_k^{\mathrm{QoS}}
      - \tau_k^{\mathrm{QoS}}\sqrt{2\ln(1/\delta_k)}
      \ge 0,
      \label{eq:QoS_bernstein_det_a} \\
    & \left\|
        \mathbf{C}_k^{1/2}
        \mathbf{A}_k^{\mathrm{QoS}}
        \mathbf{C}_k^{1/2}
      \right\|_{\!F}
      \le \tau_k^{\mathrm{QoS}},
      \label{eq:QoS_bernstein_det_b} \\
    & \left\|
        \mathbf{C}_k^{1/2}
        \mathbf{b}_k^{\mathrm{QoS}}
      \right\|_2
      \le \frac{\tau_k^{\mathrm{QoS}}}{\sqrt{2}}.
      \label{eq:QoS_bernstein_det_c}
\end{align}
\end{subequations}

The constraints in \eqref{eq:QoS_bernstein_det} are jointly convex in the optimization variables for fixed covariance matrix $\mathbf{C}_k$ and violation probability $\delta_k$, and can be expressed as second-order cone constraints. They therefore provide a tractable and conservative deterministic surrogate of the original QoS chance constraint.

\subsection{Deterministic Secrecy Chance Constraints}
We next consider the secrecy chance constraints in \eqref{eq:P1_secrecy_chance}. For a given user $k$, the secrecy-rate requirement $R_k^{\mathrm{sec}} \ge R_k^{\mathrm{sec,min}}$ can be expressed as
\begin{equation}
    \log_2\!\left(1+\gamma_k^{(\ell)}\right)
    -
    \log_2\!\left(1+\gamma_k^{(e)}\right)
    \ge R_k^{\mathrm{sec,min}},
    \label{eq:secrecy_event_log}
\end{equation}
which couples the legitimate-user SINR and the aggregate eavesdropper SINR in a nonconvex manner.

A common approach to handle \eqref{eq:secrecy_event_log} is to enforce it via a pair of sufficient SINR constraints of the form
\begin{equation}
    \gamma_k^{(\ell)} \ge \gamma_k^{\mathrm{sec},\ell},\qquad
    \gamma_k^{(e)} \le \gamma_k^{\mathrm{sec},e},
    \label{eq:secrecy_thresholds}
\end{equation}
where the thresholds $\gamma_k^{\mathrm{sec},\ell}$ and $\gamma_k^{\mathrm{sec},e}$ are selected such that \eqref{eq:secrecy_thresholds} constitutes a tight inner approximation of \eqref{eq:secrecy_event_log}. These thresholds depend on $R_k^{\mathrm{sec,min}}$ and can be obtained, for instance, via supporting-hyperplane approximations or a one-dimensional bisection search.

The legitimate-side constraint in \eqref{eq:secrecy_thresholds} has the same structure as the QoS SINR constraint and can therefore be handled identically to \eqref{eq:QoS_bernstein_det}, with $\gamma_k^{\mathrm{QoS}}$ replaced by $\gamma_k^{\mathrm{sec},\ell}$. The eavesdropper-side constraint $\gamma_k^{(e)} \le \gamma_k^{\mathrm{sec},e}$ can be equivalently rewritten as
\begin{equation}
    \gamma_k^{(e)} - \gamma_k^{\mathrm{sec},e} \le 0.
    \label{eq:eavesdropper_gamma_constraint}
\end{equation}

By substituting the aggregate wiretap SINR expression in \eqref{eq:colluding_sinr} into \eqref{eq:eavesdropper_gamma_constraint} and rearranging terms, the eavesdropper-side constraint can be reformulated as a quadratic inequality in the stacked channel-error vector of all colluding eavesdroppers. Specifically, we obtain
\begin{equation}
    \Delta\mathbf{h}_k^{(e)H}
    \mathbf{A}_k^{\mathrm{sec}}
    \Delta\mathbf{h}_k^{(e)}
    +
    2\Re\!\left\{
        \mathbf{b}_k^{\mathrm{sec}H} \Delta\mathbf{h}_k^{(e)}
    \right\}
    +
    c_k^{\mathrm{sec}}
    \;\ge\; 0,
    \label{eq:quadratic_secrecy_form}
\end{equation}
where $\Delta\mathbf{h}_k^{(e)}$ denotes the stacked channel-estimation error vector associated with all eavesdroppers colluding against user $k$, and $\mathbf{C}_k^{(e)}$ denotes the corresponding block-diagonal covariance matrix.

The secrecy chance constraint \eqref{eq:P1_secrecy_chance} can therefore be conservatively guaranteed by imposing the pair of chance constraints in \eqref{eq:secrecy_thresholds} and applying Lemma~1 separately to the legitimate-side and eavesdropper-side conditions. To this end, we introduce violation probabilities $\epsilon_k^{(\ell)}$ and $\epsilon_k^{(e)}$ such that $\epsilon_k^{(\ell)} + \epsilon_k^{(e)} \le \epsilon_k$. This yields the following sufficient deterministic conditions.

\emph{Legitimate-side secrecy constraint:}
\begin{subequations}
\label{eq:sec_legit_bernstein_det}
\begin{align}
    & \mathrm{tr}\!\left(
        \mathbf{A}_k^{\mathrm{sec},\ell}\mathbf{C}_k
      \right)
      + c_k^{\mathrm{sec},\ell}
      - \tau_k^{\mathrm{sec},\ell}\sqrt{2\ln(1/\epsilon_k^{(\ell)})}
      \ge 0,
      \\
    & \left\|
        \mathbf{C}_k^{1/2}
        \mathbf{A}_k^{\mathrm{sec},\ell}
        \mathbf{C}_k^{1/2}
      \right\|_{\!F}
      \le \tau_k^{\mathrm{sec},\ell},
      \\
    & \left\|
        \mathbf{C}_k^{1/2}
        \mathbf{b}_k^{\mathrm{sec},\ell}
      \right\|_2
      \le
      \frac{\tau_k^{\mathrm{sec},\ell}}{\sqrt{2}},
\end{align}
\end{subequations}

\emph{Eavesdropper-side secrecy constraint:}
\begin{subequations}
\label{eq:sec_eve_bernstein_det}
\begin{align}
    & \mathrm{tr}\!\left(
        \mathbf{A}_k^{\mathrm{sec},e}\mathbf{C}_k^{(e)}
      \right)
      + c_k^{\mathrm{sec},e}
      - \tau_k^{\mathrm{sec},e}\sqrt{2\ln(1/\epsilon_k^{(e)})}
      \ge 0,
      \\
    & \left\|
        \mathbf{C}_k^{(e)1/2}
        \mathbf{A}_k^{\mathrm{sec},e}
        \mathbf{C}_k^{(e)1/2}
      \right\|_{\!F}
      \le \tau_k^{\mathrm{sec},e},
      \\
    & \left\|
        \mathbf{C}_k^{(e)1/2}
        \mathbf{b}_k^{\mathrm{sec},e}
      \right\|_2
      \le
      \frac{\tau_k^{\mathrm{sec},e}}{\sqrt{2}},
\end{align}
\end{subequations}
where $\tau_k^{\mathrm{sec},\ell}\ge 0$ and $\tau_k^{\mathrm{sec},e}\ge 0$ are slack variables. The matrices and vectors $(\mathbf{A}_k^{\mathrm{sec},\ell}, \mathbf{b}_k^{\mathrm{sec},\ell}, c_k^{\mathrm{sec},\ell})$ and $(\mathbf{A}_k^{\mathrm{sec},e}, \mathbf{b}_k^{\mathrm{sec},e}, c_k^{\mathrm{sec},e})$ are obtained by rewriting the thresholds in \eqref{eq:secrecy_thresholds} into quadratic form, analogously to \eqref{eq:A_k_QoS_def}--\eqref{eq:c_k_QoS_def}.

\subsection{Deterministically Approximated Problem}
By replacing the QoS chance constraints in \eqref{eq:P1_QoS_chance} with their Bernstein-type deterministic approximations in \eqref{eq:QoS_bernstein_det}, and replacing the secrecy chance constraints in \eqref{eq:P1_secrecy_chance} with the deterministic conditions in \eqref{eq:sec_legit_bernstein_det}--\eqref{eq:sec_eve_bernstein_det}, the original stochastic optimization problem~(P1) is conservatively approximated by the following deterministic problem:
\begin{subequations}
\label{eq:ProbP1_DR}
\begin{align}
    \text{(P1-DR)}\quad
    \min_{\boldsymbol{\Xi},\{\tau\}} \quad
    & \Phi(\boldsymbol{\Xi})
      \label{eq:P1_DR_obj} \\[1mm]
    \text{s.t.}\quad
    & \eqref{eq:P1_power_constraint},\ 
      \eqref{eq:P1_RIS_constraints},\
      \eqref{eq:P1_UAV_constraint},\
      \eqref{eq:P1_error_constraint},
      \label{eq:P1_DR_inherited} \\[1mm]
    & \eqref{eq:QoS_bernstein_det}
      \quad \forall k\in\mathcal{K},
      \label{eq:P1_DR_QoS_det} \\[1mm]
    & \eqref{eq:sec_legit_bernstein_det},\
      \eqref{eq:sec_eve_bernstein_det}
      \quad \forall k\in\mathcal{K}.
      \label{eq:P1_DR_sec_det}
\end{align}
\end{subequations}
Here, $\{\tau\}$ collects all slack variables introduced by the Bernstein-type approximations.

Problem~(P1-DR) is a deterministic but still non-convex optimization problem, due to the multiplicative coupling among the BS beamforming vectors, the configurations of the UAV-RIS, STAR-RIS, and H-RIS, and the UAV placement variables. Nevertheless, all probabilistic secrecy and QoS constraints have been transformed into convex deterministic constraints for fixed subsets of optimization variables. This block-wise convex structure enables the efficient application of an alternating optimization framework combined with successive convex approximation (SCA), which will be developed in Section~V.

\section{Proposed SCA--AO Algorithm}
\label{sec:proposed_algorithm}
The distributionally robust problem~(P1-DR) in \eqref{eq:ProbP1_DR} remains non-convex due to several sources of coupling and nonlinearity. Specifically, (i) the BS beamforming vectors, the configurations of the UAV-RIS, STAR-RIS, and H-RIS, and the UAV position appear multiplicatively in the effective channels; (ii) the matrices involved in the deterministic constraints \eqref{eq:QoS_bernstein_det}--\eqref{eq:sec_eve_bernstein_det} are nonlinear functions of the optimization variables $\boldsymbol{\Xi}$; and (iii) the weighted secrecy-outage cost $\Phi(\boldsymbol{\Xi})$ relies on a smooth penalty approximation of the outage indicator functions, which is itself non-convex.  

To efficiently handle these challenges, we develop an SCA framework embedded within an alternating optimization (AO) procedure. This approach exploits the block-wise structure of~(P1-DR) to iteratively solve a sequence of convex subproblems and guarantees convergence to a stationary point of the deterministically approximated problem.

\subsection{Block-Coordinate Structure}
To exploit the structure of problem~(P1-DR), we partition the optimization variables $\boldsymbol{\Xi}$ into three disjoint blocks:
\begin{equation}
    \boldsymbol{\Xi}
    = 
    \big(
        \boldsymbol{\Xi}_{\mathrm{BF}},
        \boldsymbol{\Xi}_{\mathrm{RIS}},
        \boldsymbol{\Xi}_{\mathrm{UAV}}
      \big),
\end{equation}
where $\boldsymbol{\Xi}_{\mathrm{BF}} = \{\mathbf{w}_k\}_{k\in\mathcal{K}}$ collects the BS beamforming vectors, $\boldsymbol{\Xi}_{\mathrm{RIS}}= \{\boldsymbol{\Theta}_U,\boldsymbol{\Theta}_T,\boldsymbol{\Theta}_R,\boldsymbol{\Theta}_H\}$ denotes the configuration variables of the UAV-RIS, STAR-RIS, and H-RIS, and $\boldsymbol{\Xi}_{\mathrm{UAV}} = \mathbf{p}_U$ represents the UAV position.

For fixed $(\boldsymbol{\Xi}_{\mathrm{RIS}}, \boldsymbol{\Xi}_{\mathrm{UAV}})$, problem~(P1-DR) becomes convex in $\boldsymbol{\Xi}_{\mathrm{BF}}$ after applying SCA to linearize the non-convex quadratic terms arising from the effective channels. Similarly, for fixed $(\boldsymbol{\Xi}_{\mathrm{BF}}, \boldsymbol{\Xi}_{\mathrm{UAV}})$, the RIS/STAR-RIS/H-RIS subproblem can be transformed into a convex program by handling the unit-modulus and amplitude constraints through standard convex envelopes or difference-of-convex (DC) decompositions. Finally, for fixed $(\boldsymbol{\Xi}_{\mathrm{BF}}, \boldsymbol{\Xi}_{\mathrm{RIS}})$, the optimization of the UAV position $\mathbf{p}_U$ becomes convex after SCA linearization of the distance-dependent pathloss terms $\beta_{a,b}(d_{a,b}(\mathbf{p}_U))$.

This block-wise convexity property naturally motivates an AO scheme, in which the beamforming, reconfigurable surface configurations, and UAV position are updated cyclically while keeping the remaining blocks fixed.

\subsection{Successive Convex Approximation (SCA)}
At iteration $t$, let $\boldsymbol{\Xi}^{(t)}$ denote the current feasible solution. The key principle of SCA is to replace each non-convex function $f(\boldsymbol{\Xi})$ with a convex surrogate $\widetilde{f}(\boldsymbol{\Xi};\boldsymbol{\Xi}^{(t)})$ constructed around $\boldsymbol{\Xi}^{(t)}$ such that
\begin{enumerate}
    \item $\widetilde{f}(\boldsymbol{\Xi}^{(t)};\boldsymbol{\Xi}^{(t)}) = f(\boldsymbol{\Xi}^{(t)})$ (value consistency),
    \item $\widetilde{f}(\boldsymbol{\Xi};\boldsymbol{\Xi}^{(t)}) \ge f(\boldsymbol{\Xi})$ for all $\boldsymbol{\Xi}$ (global upper bound),
    \item $\widetilde{f}(\boldsymbol{\Xi};\boldsymbol{\Xi}^{(t)})$ is convex in $\boldsymbol{\Xi}$.
\end{enumerate}
These conditions ensure feasibility preservation and monotonic descent of the objective value across SCA iterations.

For quadratic terms of the form $|\mathbf{h}^H\mathbf{w}_i|^2$ appearing in $\mathbf{A}_k^{\mathrm{QoS}}$ and $\mathbf{A}_k^{\mathrm{sec}}$, we use first-order linearization. Specifically,
\begin{equation}
    |\mathbf{h}^H \mathbf{w}|^2
    =
    \mathbf{w}^H \mathbf{H} \mathbf{w},
\end{equation}
where $\mathbf{H}=\mathbf{h}\mathbf{h}^H$. Given a reference point $\mathbf{w}^{(t)}$, a convex upper bound is obtained as
\begin{equation}
    \mathbf{w}^H \mathbf{H}\mathbf{w}
    \le
    \mathbf{w}^{(t)H}\mathbf{H}\mathbf{w}^{(t)}
    +
    2\Re\!\big\{(\mathbf{H}\mathbf{w}^{(t)})^H(\mathbf{w}-\mathbf{w}^{(t)})\big\},
\label{eq:SCA_linearization}
\end{equation}
which is tight at $\mathbf{w}=\mathbf{w}^{(t)}$.

For the unit-modulus constraints associated with RIS phase shifts, i.e., $|\theta_m|=1$, we adopt a standard DC representation
\begin{equation}
    |\theta_m|^2 = 1
    \quad\Longleftrightarrow\quad
    |\theta_m|^2 - 1 = 0,
\end{equation}
and linearize the concave component around the current iterate $\theta_m^{(t)}$ using a first-order Taylor expansion, yielding a convex inner approximation.

Finally, the UAV position $\mathbf{p}_U$ influences the effective channels through the distance-dependent large-scale fading terms $\beta_{a,b}(d_{a,b}(\mathbf{p}_U))$. These terms are convexified by first-order linearization with respect to $\mathbf{p}_U$ around the current point $\mathbf{p}_U^{(t)}$. Collectively, these surrogate constructions yield a sequence of convex subproblems within each AO block.

\subsection{Overall SCA--AO Procedure}
At iteration $t$, given the current feasible point $\boldsymbol{\Xi}^{(t)}$, the proposed algorithm updates the optimization variables by cyclically optimizing each block while keeping the remaining blocks fixed.

\paragraph*{1) Beamforming Update}
With $(\boldsymbol{\Xi}_{\mathrm{RIS}},\boldsymbol{\Xi}_{\mathrm{UAV}})$ fixed at their current values, we solve the following convex subproblem:
\begin{equation}
\label{eq:BF_subproblem}
\begin{aligned}
    \min_{\boldsymbol{\Xi}_{\mathrm{BF}},\{\tau\}} \quad
        & \widetilde{\Phi}\big(
            \boldsymbol{\Xi}_{\mathrm{BF}};
            \boldsymbol{\Xi}^{(t)}
        \big) \\
    \text{s.t.}\quad
        & \eqref{eq:P1_power_constraint},\ 
          \eqref{eq:QoS_bernstein_det},\ 
          \eqref{eq:sec_legit_bernstein_det},\
          \eqref{eq:sec_eve_bernstein_det},
\end{aligned}
\end{equation}
where $\widetilde{\Phi}$ denotes a convex upper bound of the weighted secrecy-outage cost obtained via SCA around $\boldsymbol{\Xi}^{(t)}$.

\paragraph*{2) RIS / STAR-RIS / H-RIS Update}
With $(\boldsymbol{\Xi}_{\mathrm{BF}},\boldsymbol{\Xi}_{\mathrm{UAV}})$ fixed, the RIS-related variables are updated by solving the convexified subproblem
\begin{equation}
\label{eq:RIS_subproblem}
\begin{aligned}
    \min_{\boldsymbol{\Xi}_{\mathrm{RIS}},\{\tau\}} \quad
        & \widetilde{\Phi}\big(
            \boldsymbol{\Xi}_{\mathrm{RIS}};
            \boldsymbol{\Xi}^{(t)}
        \big) \\
    \text{s.t.}\quad
        & \eqref{eq:P1_RIS_constraints},\ 
          \eqref{eq:QoS_bernstein_det},\ 
          \eqref{eq:sec_legit_bernstein_det},\ 
          \eqref{eq:sec_eve_bernstein_det},
\end{aligned}
\end{equation}
where all non-convex terms and unit-modulus constraints are handled using SCA and DC-based convex approximations.

\paragraph*{3) UAV Position Update}
With $(\boldsymbol{\Xi}_{\mathrm{BF}},\boldsymbol{\Xi}_{\mathrm{RIS}})$ fixed, the UAV position is updated by solving
\begin{equation}
\label{eq:UAV_subproblem}
\begin{aligned}
    \min_{\mathbf{p}_U,\{\tau\}} \quad
        & \widetilde{\Phi}\big(
            \mathbf{p}_U;
            \boldsymbol{\Xi}^{(t)}
        \big) \\
    \text{s.t.}\quad
        & \mathbf{p}_U \in \mathcal{P}_U,\ 
          \eqref{eq:QoS_bernstein_det},\
          \eqref{eq:sec_legit_bernstein_det},\
          \eqref{eq:sec_eve_bernstein_det},
\end{aligned}
\end{equation}
where the distance-dependent pathloss terms are linearized around the current position $\mathbf{p}_U^{(t)}$.

After completing the above three updates, the algorithm yields a new iterate $\boldsymbol{\Xi}^{(t+1)}$, and the procedure is repeated until convergence.

\subsection{Convergence Guarantee}
The convex surrogate functions constructed in the SCA steps satisfy the standard majorization properties, namely tightness at the current iterate and global upper-bounding of the original non-convex functions. As a result, each block update yields a non-increasing value of the original objective function, i.e.,
\begin{equation}
    \Phi(\boldsymbol{\Xi}^{(t+1)}) \le \Phi(\boldsymbol{\Xi}^{(t)}).
\end{equation}
Since the weighted secrecy-outage cost is lower bounded and each convex subproblem is solved optimally, the sequence $\{\boldsymbol{\Xi}^{(t)}\}$ generated by the proposed SCA--AO algorithm is guaranteed to converge.

Moreover, following standard results in successive convex approximation theory~\cite{ref:SCA}, any limit point of the sequence $\{\boldsymbol{\Xi}^{(t)}\}$ satisfies the Karush--Kuhn--Tucker (KKT) conditions of the deterministically approximated problem~(P1-DR), and is therefore a stationary point of~(P1-DR).

\begin{algorithm}[!ht]
\caption{SCA--AO Algorithm for Solving (P1-DR)}
\label{alg:SCA_AO}
\begin{algorithmic}[1]
\State \textbf{Input:} Channel estimates $\{\widehat{\mathbf{h}}_u\}$, covariance
matrices $\{\mathbf{C}_u\}$, RIS/H-RIS/STAR-RIS feasible sets, UAV region
$\mathcal{P}_U$, outage weights $\{\omega_k\}$, tolerance $\varepsilon$.
\State \textbf{Initialize:} Feasible $\boldsymbol{\Xi}^{(0)}$ (e.g., MRT/ZF
beamforming, random RIS phases, UAV hovering location).
\State \textbf{Repeat} for $t=0,1,2,\ldots$
    \State \quad \textbf{Beamforming update:} Solve \eqref{eq:BF_subproblem} via convex
    optimization to obtain $\boldsymbol{\Xi}_{\mathrm{BF}}^{(t+1)}$.
    \State \quad \textbf{RIS/STAR-RIS/H-RIS update:} Solve \eqref{eq:RIS_subproblem}
    to obtain $\boldsymbol{\Xi}_{\mathrm{RIS}}^{(t+1)}$.
    \State \quad \textbf{UAV position update:} Solve \eqref{eq:UAV_subproblem} to
    obtain $\mathbf{p}_U^{(t+1)}$.
    \State \quad \textbf{Update:} Form $\boldsymbol{\Xi}^{(t+1)}$.
\State \textbf{Until} 
    $|\Phi(\boldsymbol{\Xi}^{(t+1)}) - \Phi(\boldsymbol{\Xi}^{(t)})| \le \varepsilon$.
\State \textbf{Output:} Stationary solution $\boldsymbol{\Xi}^{\star}$.
\end{algorithmic}
\end{algorithm}

\subsection{Remark on Conservativeness and Tuning of Bernstein Bounds}
The Bernstein-type deterministic approximations employed in
\eqref{eq:QoS_bernstein_det}--\eqref{eq:sec_eve_bernstein_det} are inherently conservative, in the sense that satisfaction of the resulting deterministic constraints guarantees the original chance constraints, but does not ensure tight equivalence. The level of conservativeness depends on several factors, including:
\begin{enumerate}
    \item the prescribed violation probabilities $\delta_k$, $\epsilon_k^{(\ell)}$, and $\epsilon_k^{(e)}$;
    \item the eigenvalue distribution and conditioning of the channel-error covariance matrices $\mathbf{C}_k$ and $\mathbf{C}_k^{(e)}$;
    \item the tightness of the successive convex approximation (SCA) surrogates used to upper-bound the quadratic forms;
    \item the degree of statistical correlation between the legitimate-user and eavesdropper channel uncertainties.
\end{enumerate}

To strike a practical balance between robustness and performance, several tuning strategies can be adopted. First, a simple and effective choice is to split the secrecy violation probability evenly as $\epsilon_k^{(\ell)} = \epsilon_k^{(e)} = \epsilon_k/2$, which avoids biasing the design toward either the legitimate or eavesdropper-side constraint. Second, a continuation strategy can be employed in which the violation probabilities $\delta_k$ and $\epsilon_k$ are initialized at relatively loose values and gradually tightened across SCA iterations, thereby mitigating excessive conservativeness in early iterations. Finally, the achieved outage probabilities can be validated empirically via Monte-Carlo simulations to ensure consistency with the target reliability levels. If the resulting design is found to be overly conservative, the violation probabilities may be slightly relaxed or tighter SCA linearizations may be adopted.

\section{Computational Complexity Analysis}
\label{sec:complexity}
In this section, we analyze the computational complexity of the proposed SCA-AO algorithm, which is summarized in Algorithm~\ref{alg:SCA_AO}. At each iteration, the algorithm solves three convex subproblems corresponding to the beamforming update, the RIS/STAR-RIS/H-RIS configuration update, and the UAV position update. All subproblems are solved using interior-point methods, whose worst-case complexity scales cubically with the number of optimization variables and the number of active convex constraints~\cite{ref:boyd}.

\subsection{Beamforming Update}
The beamforming subproblem optimizes $N_t|\mathcal{K}|$ complex-valued variables (equivalently $2N_t|\mathcal{K}|$ real variables). After SCA linearization, the
dominant computational burden arises from the second-order cone (SOC) constraints introduced by the Bernstein-type QoS and secrecy conditions. The resulting worst-case complexity of the beamforming update scales as
\begin{equation}
\mathcal{O}\!\left((N_t|\mathcal{K}|)^3\right).
\end{equation}

\subsection{RIS/STAR-RIS/H-RIS Update}
The RIS-related subproblem optimizes the configuration variables of four reconfigurable surfaces: $M_U$ UAV-RIS elements, $M_S$ STAR-RIS elements (each with transmit and reflect coefficients), and $M_H$ holographic RIS elements. The total number of optimization variables can be approximated as
\begin{equation}
N_{\mathrm{RIS}} \approx M_U + 2M_S + M_H.
\end{equation}
After convexification, the dominant complexity is due to SOC constraints induced by the deterministic QoS and secrecy conditions, yielding a worst-case complexity of
\begin{equation}
\mathcal{O}\!\left(N_{\mathrm{RIS}}^3\right).
\end{equation}
In holographic RIS deployments, where $M_H \gg M_S, M_U$, this block constitutes the primary computational bottleneck.

\subsection{UAV Position Update}
The UAV position update optimizes the three-dimensional vector $\mathbf{p}_U\in\mathbb{R}^3$. After SCA linearization of the distance-dependent pathloss terms, the resulting subproblem involves only low-dimensional SOC constraints. Consequently, the computational cost of this block is negligible compared with the beamforming and RIS updates and can be regarded as constant per
iteration.

\subsection{Overall Complexity and Scaling}
Let $T_{\mathrm{iter}}$ denote the number of AO iterations required for convergence. The total worst-case computational complexity of the proposed SCA-AO algorithm scales as
\begin{equation}
\mathcal{O}\!\left(
T_{\mathrm{iter}}
\Big[
(N_t|\mathcal{K}|)^3
+
(M_U + 2M_S + M_H)^3
\Big]
\right),
\end{equation}
which is dominated by the beamforming and RIS/STAR-RIS/H-RIS update steps.

While the resulting cubic complexity is typical for robust convex optimization frameworks, the proposed algorithm is primarily intended for offline planning
and configuration. In fast-varying scenarios, lightweight learning-based surrogates can be trained using the SCA--AO solutions to enable real-time
adaptation with negligible online computational cost.

\section{Simulation Setup}
\label{sec:simulation_setup}
We now describe the simulation environment used to evaluate the performance of the proposed hybrid UAV--RIS / STAR-RIS / H-RIS system. All channel realizations are generated in accordance with the relevant 3GPP and ITU-R specifications. Unless otherwise stated, the proposed framework is instantiated at a representative mmWave carrier frequency of $28$~GHz, and the corresponding frequency-dependent pathloss, blockage behavior, and outdoor-to-indoor penetration losses are applied following 3GPP TR~38.901, TR~36.873, and ITU-R P.2109. The key simulation parameters, including carrier frequency, bandwidth, antenna configurations, node densities, mobility constraints, and hardware impairment factors, are summarized in Table~\ref{tab:sim_parameters}. All reported performance metrics are obtained by averaging over $10^4$ independent channel realizations, which include random user and eavesdropper locations, small-scale fading, blockage states, and CSI estimation errors, thereby ensuring statistical reliability.

Residual hardware impairments at the BS and user terminals are modeled using the widely adopted error vector magnitude (EVM) framework. The transmitter-side distortion at the BS is characterized by an EVM level of $\kappa_t=-28~\mathrm{dB}$, while the receiver-side distortion at each user
terminal is characterized by $\kappa_{r,u}=-30~\mathrm{dB}$, in accordance with the transceiver requirements specified in 3GPP TR~38.104. These impairments are modeled as additive distortion noises whose variances scale with the instantaneous signal power. Unless otherwise stated, the phase responses of the RIS, STAR-RIS, and H-RIS are quantized with $B$-bit resolution to reflect practical hardware limitations in reconfigurable surface implementations.

Eavesdroppers are spatially distributed according to a homogeneous PPP with density $\lambda_{\mathrm{eve}}$. All eavesdroppers are assumed to be fully colluding, such that their received signals can be jointly processed, effectively forming a virtual multi-antenna wiretap receiver. In addition to passive interception, a subset of eavesdroppers may actively transmit jamming signals with power $P_{\mathrm{jam}}$. Both passive and active eavesdroppers follow the same 3GPP-compliant channel models as legitimate users, including mobility and blockage effects, thereby representing a worst-case adversarial scenario.

\begin{table}[!t]
\centering
\caption{Simulation Parameters}
\label{tab:sim_parameters}
\renewcommand{\arraystretch}{1.15}
\begin{tabular}{p{4.4cm} p{3.4cm}}
\toprule
Parameter & Value \\
\midrule
Carrier frequency & 28~GHz \\
System bandwidth & 5~GHz \\
BS antenna array & $N_t = 16$ \\
BS height & $h_{\mathrm{BS}} = 25$~m \\
UAV altitude & $h_{\mathrm{UAV}} = 80$~m \\
UAV horizontal region & $x_U,y_U \in [-150,150]$~m \\
UAV mobility constraint & $v_{\mathrm{UAV}} \le 15$~m/s \\
UAV-RIS size & $M_U = 80$ \\
STAR-RIS size & $M_S = 256$ \\
H-RIS size& $M_H = 1024$\\
RIS phase quantization & $B = 3$ bits \\
Shadowing & $\sigma_{\mathrm{LoS}} = 4$ dB, $\sigma_{\mathrm{NLoS}} = 7$ dB \\
Thermal noise power & $-174$~dBm/Hz \\
BS transmit power & $P_{\max} = 30$~dBm \\
Jamming power (active eaves.)& $P_{\mathrm{jam}} = 10$~dBm \\
Hardware impairment factors & $\kappa_t = -28$~dB, $\kappa_{r,u} = -30$~dB \\
Target secrecy rate & $R_k^{\mathrm{sec,min}} = 0.5$ bps/Hz \\
Target QoS rate & $R_k^{\mathrm{QoS}} = 1$ bps/Hz \\
Secrecy violation probability & $\epsilon_k = 0.05$ \\
QoS violation probability & $\delta_k = 0.05$ \\
\bottomrule
\end{tabular}
\end{table}

\begin{figure}[htbp]
\centering
\includegraphics[width=3.2in]{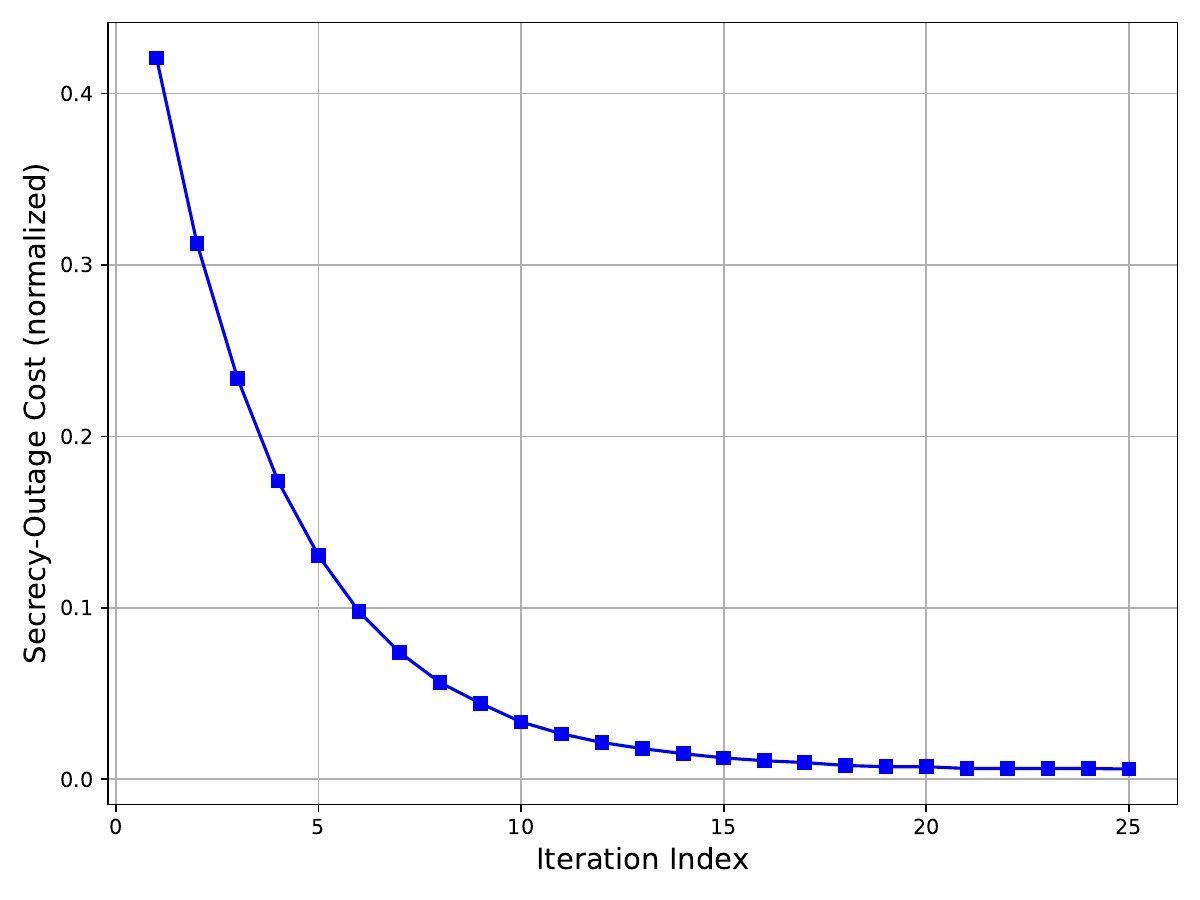}
\caption{Convergence.}
\label{fig:conv}
\end{figure}

The convergence behavior of the proposed algorithm is illustrated in Fig.~\ref{fig:conv}. The normalized secrecy-outage cost decreases monotonically with the iteration index and converges within approximately $15$--$20$ iterations. This confirms that the deterministic reformulation and alternating optimization procedure lead to a stable solution with fast convergence, rendering the proposed algorithm suitable for practical implementation.

\begin{figure}[htbp]
\centering
\includegraphics[width=3.2in]{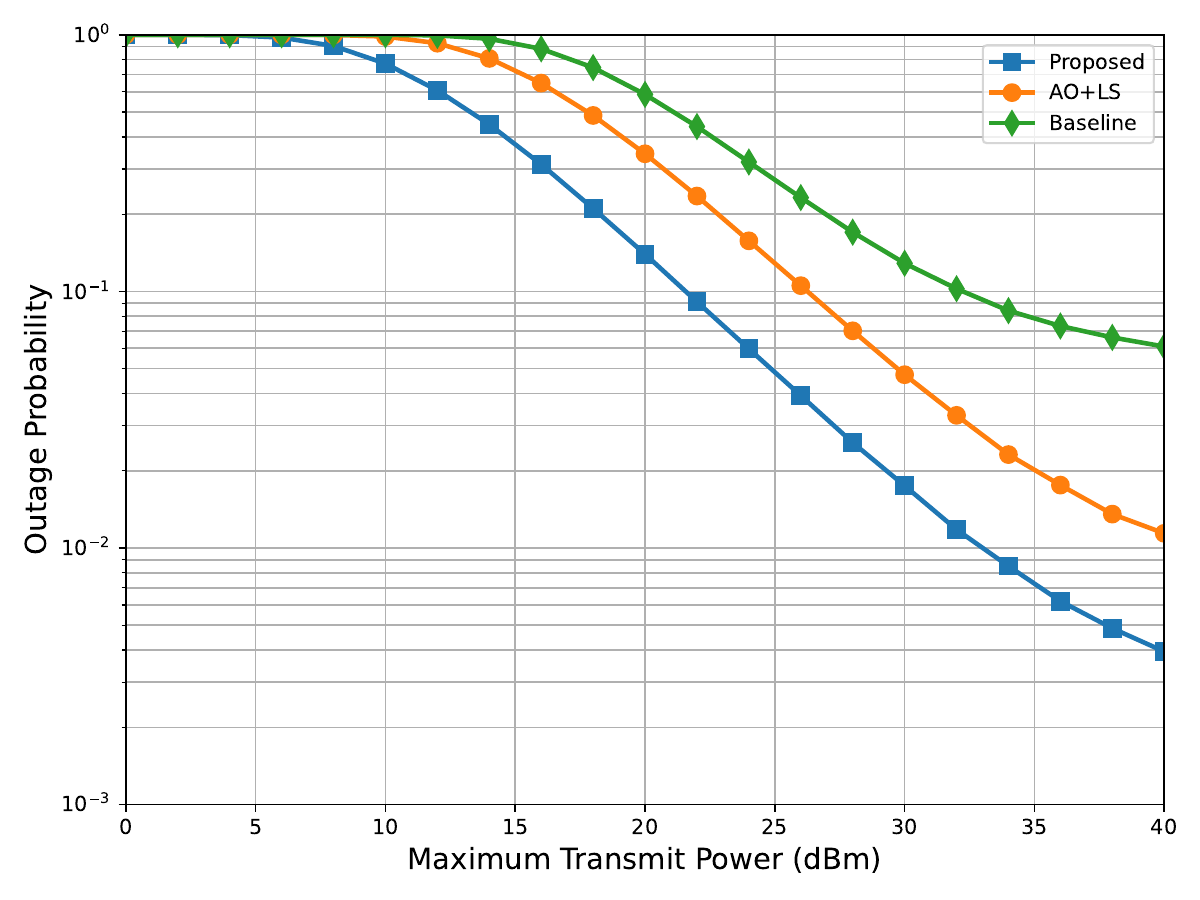}
\caption{Weighted secrecy outage probability versus BS transmit power.}
\label{fig:fig2}
\end{figure}

Fig.~\ref{fig:fig2} illustrates the secrecy outage probability as a function of the maximum transmit power for the proposed scheme, AO+LS, and the baseline. As the transmit power increases from $0$ to $40$~dBm, the secrecy outage probability decreases for all schemes due to the improved received signal quality at the legitimate users.
At low transmit power (e.g., below $10$~dBm), all schemes exhibit outage probabilities close to unity, indicating that the secrecy constraints cannot be reliably satisfied in the noise-limited regime. As the transmit power increases, the performance gap among the schemes becomes evident. For instance, at $30$~dBm, the proposed scheme achieves a secrecy outage probability on the order of $10^{-2}$, whereas AO+LS and the baseline attain outage probabilities of approximately $5\times10^{-2}$ and $10^{-1}$, respectively. This corresponds to roughly one order-of-magnitude improvement over the baseline and a factor-of-five reduction compared with AO+LS.
Moreover, while the baseline exhibits a clear outage floor in the high-power regime, the proposed scheme continues to improve with increasing transmit power, indicating that the additional power is more effectively exploited to enhance secrecy rather than being equally beneficial to colluding eavesdroppers.

\begin{figure}[htbp]
\centering
\includegraphics[width=3.2in]{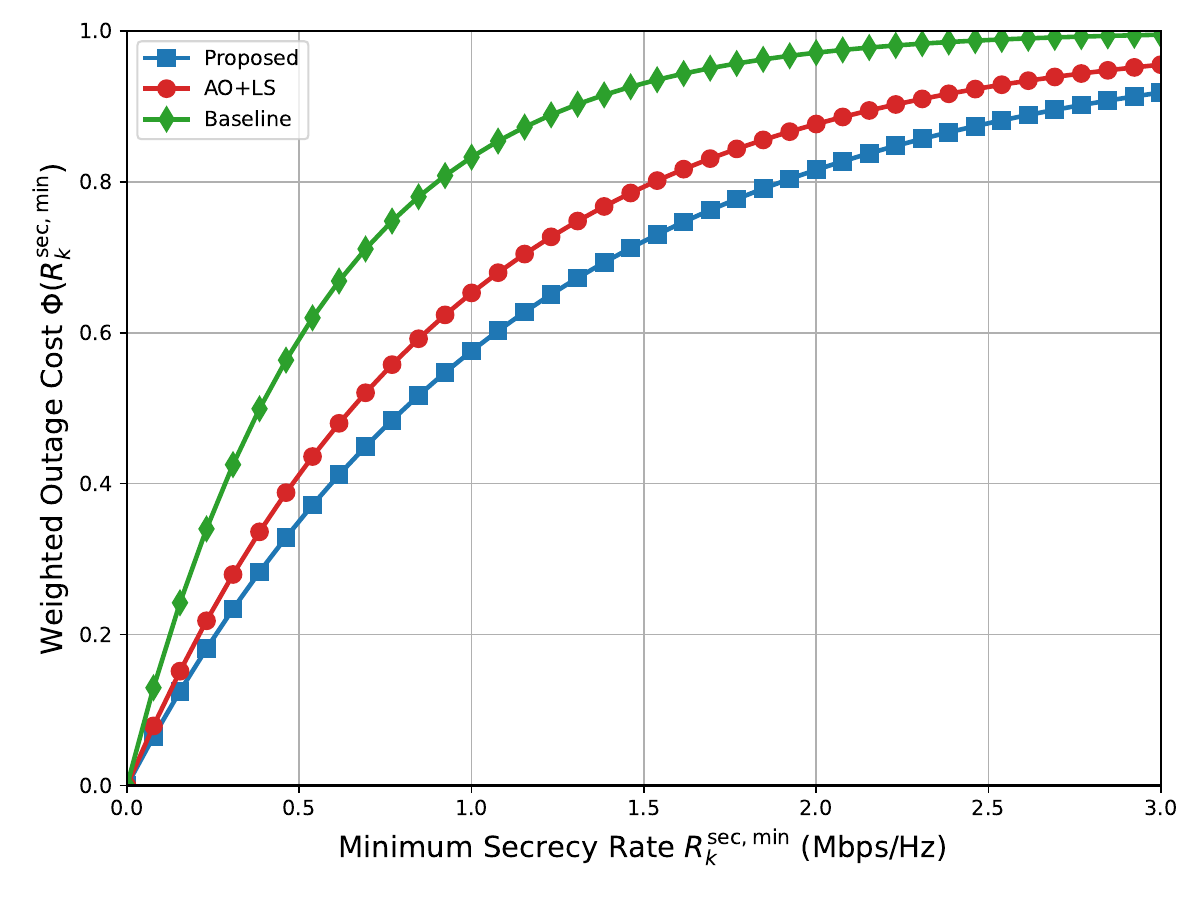}
\caption{Weighted secrecy outage probability versus the minimum secrecy rate $R_k^{\mathrm{sec,min}}$.}
\label{fig:fig3}
\end{figure}

Fig.~\ref{fig:fig3} depicts the weighted outage cost as a function of the minimum secrecy rate requirement $R_k^{\mathrm{sec,min}}$. For all schemes, the outage cost increases monotonically with $R_k^{\mathrm{sec,min}}$, since more stringent secrecy requirements reduce the feasible solution space.
Quantitatively, at a moderate secrecy requirement of $R_k^{\mathrm{sec,min}} = 1$~Mbps/Hz, the proposed scheme incurs a weighted outage cost of approximately $0.6$, compared with about $0.7$ for AO+LS and $0.85$ for the baseline. As $R_k^{\mathrm{sec,min}}$ increases to $2$~Mbps/Hz, the performance gap further widens: the outage cost of the baseline approaches unity, while the proposed scheme remains below $0.8$. These results indicate that the proposed design can sustain higher secrecy requirements with substantially lower outage penalties.

\begin{figure}[htbp]
\centering
\includegraphics[width=3.2in]{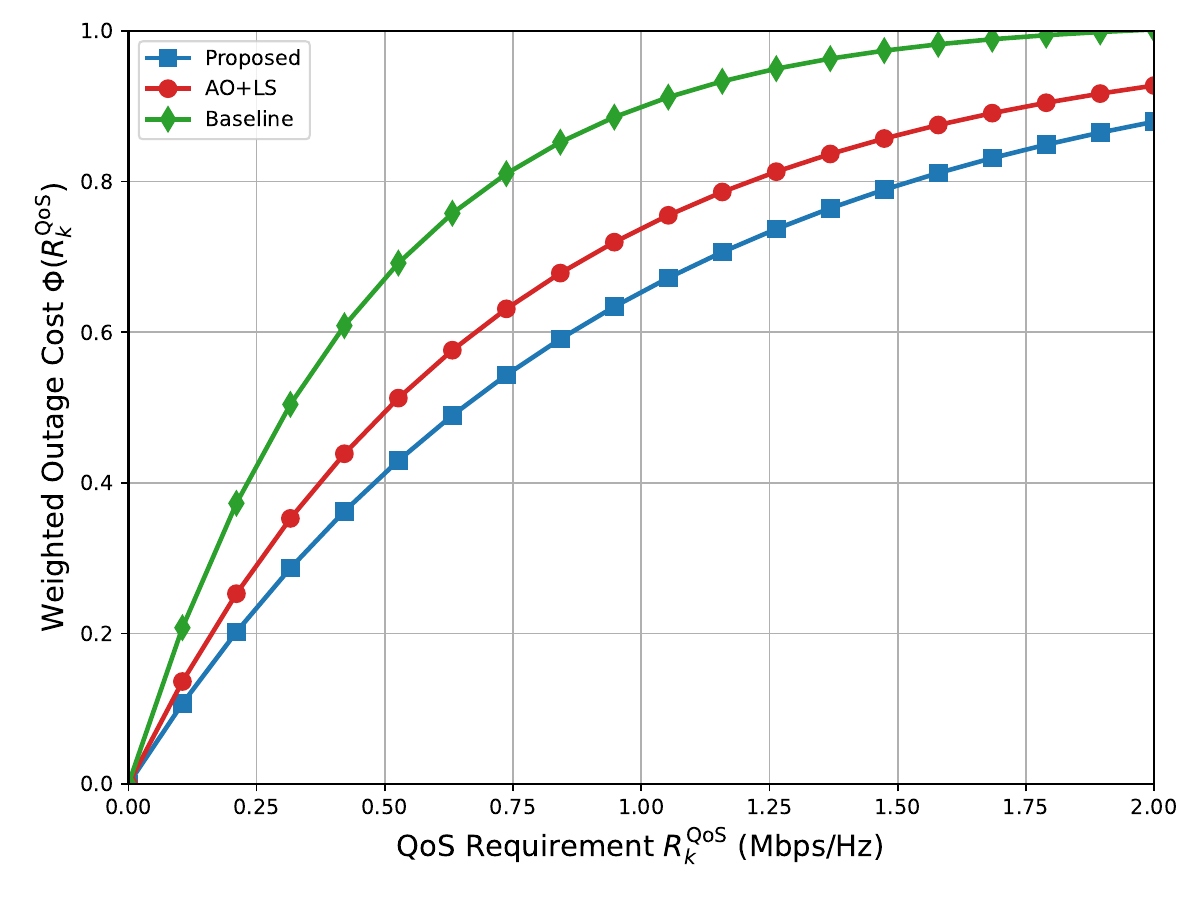}
\caption{Weighted secrecy outage probability versus the QoS requirement $ R_k^{\mathrm{QoS}}$.}
\label{fig:fig4}
\end{figure}

Fig.~\ref{fig:fig4} shows the weighted outage cost as a function of the minimum QoS requirement. As the QoS requirement increases, the outage cost rises for all schemes due to intensified competition for limited transmission resources.
For example, at a QoS requirement of $R_k^{\mathrm{QoS}} = 1$~Mbps/Hz, the proposed scheme achieves a weighted outage cost of approximately $0.65$, while AO+LS and the baseline yield outage costs of about $0.75$ and $0.9$, respectively. At higher QoS requirements (e.g., $2$~Mbps/Hz), the baseline rapidly approaches an outage cost of one, whereas the proposed scheme maintains a non-saturated performance level. This demonstrates that the proposed framework offers improved robustness to heterogeneous QoS demands under secrecy constraints.

\section{Conclusion}
This paper studied secure communication in a hybrid reconfigurable wireless architecture that integrates a UAV-mounted RIS, a STAR-RIS, and a holographic
RIS to support mixed indoor and outdoor users. By incorporating UAV mobility, indoor-outdoor propagation, hardware impairments, and statistical CSI uncertainty, a secrecy-outage minimization problem was formulated to jointly optimize base-station beamforming, multi-surface reconfigurable coefficients,
and UAV placement under secrecy and QoS constraints. Numerical results demonstrated that the proposed hybrid design achieves significantly improved secrecy outage performance compared with conventional single-surface and UAV-only baselines, while highlighting the importance of realistic modeling assumptions.

\bibliographystyle{IEEEtran}
\bibliography{references}

\end{document}